\documentclass[apl,aps,twocolumn]{revtex4-1}
\usepackage{amsmath}
\usepackage{graphicx} 
\usepackage{dcolumn} 
\usepackage{bm} 
\usepackage{dcolumn}
\usepackage{color} 
\usepackage{colortbl}
\usepackage{epstopdf}
\usepackage{amssymb}
\usepackage{verbatim}

\newcommand{\ee}[1]{\cdot10^{#1}}
\newcommand{\mr}[1]{\mathrm{#1}}
\newcommand{\unit}[1]{\,\mathrm{#1}}
\newcommand{\um}{\,\mu{\rm m}}
\newcommand{\uN}{\,\mu{\rm N}}

\newcommand{\kT}{k_{\rm B}T}
\newcommand{\kB}{k_{\rm B}}

\newcommand{\rtHz}{\sqrt{\mr{Hz}}}

\newcommand{\etal}{\textit{et al.}}

\newcommand{\fc}{f_\mr{c}}
\newcommand{\kc}{k}
\newcommand{\xrms}{x_\mr{rms}}
\newcommand{\Fmin}{F_\mr{min}}
\newcommand{\Plaser}{P_\mr{laser}}
\newcommand{\Tstage}{T_\mr{stage}}

\newcommand{\Df}{\Delta f}
\newcommand{\Fth}{F_\mr{th}}

\newcommand{\Vrms}{V_\mr{rms}}

\newcommand{\SV}{\mr{S_V}}

\newcommand{\cRT}{c_\mr{RT}}
\newcommand{\fcRT}{f_{c,\mr{RT}}}
\newcommand{\kRT}{k_\mr{RT}}
\newcommand{\sRT}{s_\mr{RT}}

\newcommand{\TRT}{T_\mr{RT}}
\newcommand{\VrmsRT}{V_\mr{rms,RT}}

\newcommand{\cmK}{c_\mr{mK}}
\newcommand{\fcmK}{f_{c,\mr{mK}}}

\newcommand{\kmK}{k_\mr{mK}}
\newcommand{\QmK}{Q_\mr{mK}}
\newcommand{\SFmK}{S_{F,\mr{mK}}}
\newcommand{\smK}{s_\mr{mK}}
\newcommand{\TmK}{T_\mr{mK}}
\newcommand{\VrmsmK}{V_\mr{rms,mK}}

\definecolor{sblue}{RGB}{179,211,231}
\definecolor{sorange}{RGB}{255,224,192}
\definecolor{sgreen}{RGB}{179,211,171}

\begin{document}

\title{Nanoladder cantilevers made from diamond and silicon}

\author{M. H\'{e}ritier$^{1}$}
\author{A. Eichler$^{1}$}
\email{eichlera@phys.ethz.ch}
\author{Y. Pan$^{2}$}
\author{U. Grob$^{1}$}
\author{I. Shorubalko$^{3}$}
\author{M. D. Krass$^{1}$}
\author{Y. Tao$^{2}$}
\email{tao@rowland.harvard.edu}
\author{C. L. Degen$^{1}$}
\email{degenc@ethz.ch}
\affiliation{$^{1}$Department of Physics, ETH Zurich, Switzerland}
\affiliation{$^{2}$Rowland Institute at Harvard, Cambridge MA, USA}
\affiliation{$^{3}$Transport at Nanoscale Interfaces, Empa, Swiss Federal Laboratories for Materials Science and Technology, Duebendorf, Switzerland}

\date{\today}

\begin{abstract}
We present a ``nanoladder'' geometry that minimizes the mechanical dissipation of ultrasensitive cantilevers.
A nanoladder cantilever consists of a lithographically patterned scaffold of rails and rungs with feature size $\sim$ 100 nm.
Compared to a rectangular beam of the same dimensions, the mass and spring constant of a nanoladder are each reduced by roughly two orders of magnitude.
We demonstrate a low force noise of $158 \substack{+62 \\ -42} \unit{zN}$ and $190 \substack{+42 \\ -33} \unit{zN}$ in a one-Hz bandwidth for devices made from silicon and diamond, respectively, measured at temperatures between 100--150 mK.
As opposed to bottom-up mechanical resonators like nanowires or nanotubes, nanoladder cantilevers can be batch-fabricated using standard lithography, which is a critical factor for applications in scanning force microscopy.
\end{abstract}

\maketitle
Nanomechanical sensors have emerged as powerful tools for the detection of minute masses~\cite{Ekinci2004,Chui2008,Naik2009,Chaste2012,Larsen2013,Gil-Santos2015} and forces~\cite{Mamin2001,Arlett2006,Teufel2009,Gavartin2012,Moser2013,Weber2015,Reinhard2016,deLepinay2017}, opening avenues into nanobiology, chemistry and solid state physics~\cite{Rugar2004,Degen2009,Bleszynski-Jayich2009,Wang2010,Hanay2012,Olcum2014,Tavernarakis2014,Nichol2012,Payne2015}. Similar to the atomic force microscope~\cite{Binnig1986}, which was one of the key enabling instruments for the rise of nanotechnology, new generations of nanomechanical sensors will be instrumental in carrying industry to a new level of miniaturization and material control.
Bottom-up fabricated nanoresonators based on individual, doubly-clamped carbon nanotubes have recently demonstrated a mass sensitivity corresponding to a single hydrogen atom and a force resolution in the zeptonewton range~\cite{Chaste2012,Moser2013}.  A similar level of sensitivity has been achieved with untethered resonators like single trapped ions~\cite{Biercuk2010} and optically levitated nanoparticles~\cite{Hempston2017}. However, the practical application of bottom-up sensors has so far been limited because they are difficult to implement in a scanning probe apparatus and are, in many cases, subject to large variations in quality, size and geometry.

Top-down fabricated nanoresonators have remained the preferred devices for sensitive force measurements.  They can be supplied in large numbers with small device-to-device variations, can take advantage of a wide range of substrate materials, and also be geometry-tailored for specific applications.  This makes top-down devices comparatively cheap and reliable.  Among them, singly-clamped cantilever beams form a particularly useful and versatile class of nanomechanical sensors due to their suitability for scanning probe experiments, such as atomic force microscopy (AFM)~\cite{Binnig1986}, magnetic force microscopy (MFM)~\cite{Rugar1990} and magnetic resonance force microscopy (MRFM)~\cite{Sidles1995,Poggio2010}.  Owing to the limited spatial resolution of top-down lithography, however, the sensitivity of cantilever beams has trailed behind that of bottom-up devices.  Current record sensitivities are on the order of $500-800\unit{zN/\rtHz}$ for devices made from low-dissipation materials including silicon \cite{Mamin2001} and diamond \cite{Tao2014}.

The sensitivity of nanomechanical detectors is ultimately limited by thermomechanical force noise.  In the absence of other noise sources, the minimum measurable force is given by $\Fth = \left(4 \gamma k_B T B\right)^{1/2}$, where $\gamma$ describes the resonator dissipation, $\kT$ is the thermal energy, and $B$ the measurement bandwidth \cite{Stowe1997}.  Apart from reducing the operation temperature, the only means for reducing the force noise is through a reduction of the resonator dissipation $\gamma = \sqrt{k m}/Q$.  This translates into (i) combining a low spring constant $k$ with (ii) a small effective mass $m$ and (iii) employing a material that supports a high mechanical quality factor $Q$.  Intuitively, both the spring constant and mass can be reduced by decreasing the cross section of the cantilever while maintaining the beam length.  

\begin{figure}
\includegraphics[width=\columnwidth]{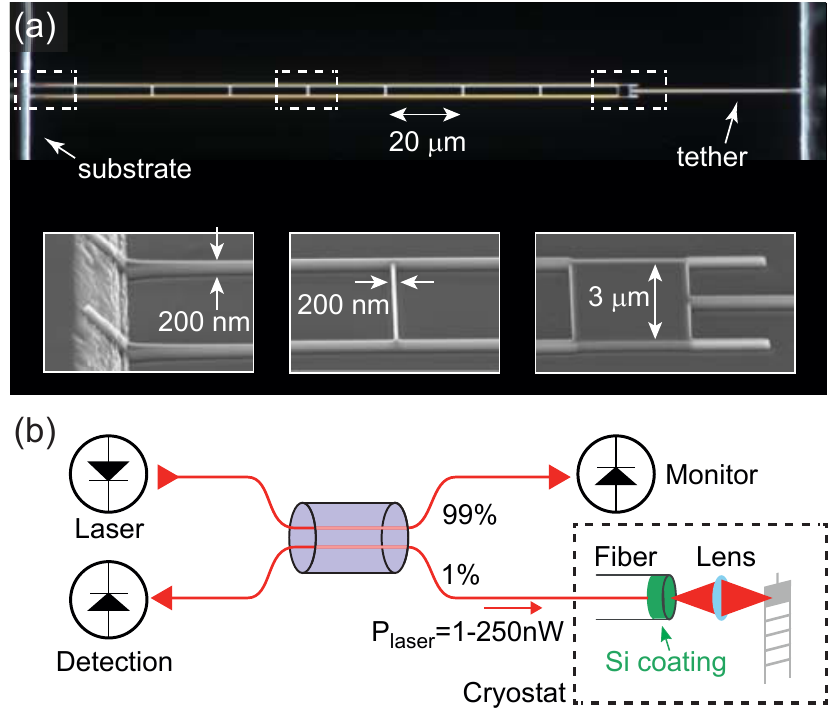}
\caption{Nanoladder design and interferometric detection. (a) Optical micrograph of a $150\unit{\um}$-long diamond nanoladder device. The nanoladder consists of two parallel nanowires (length $L$, width $w$, height $t$) that are connected by rungs every $20\unit{\um}$. Insets show magnified electron micrographs of the base, middle and tip regions of the device. A paddle located near the tip of the ladder allows for optical readout of the cantilever vibration. Devices fabricated for this study used $t \approx w = 100$ to $300\unit{nm}$ and $L=100$ to $300\unit{\um}$. (b) Schematic of the nanowatt fiber-optic interferometer used for displacement detection \cite{Rugar1989,Mamin2001}.
}
  \label{fig:Fig1}
\end{figure}

In this Letter, we introduce a ``nanoladder'' geometry to reduce the force noise of top-down, batch produced ultrasensitive cantilevers.  The nanoladder cantilever consists of two parallel wires joined by struts along their length (see Figure \ref{fig:Fig1}).  The resulting structure maximizes stability along the transverse and longitudinal directions while minimizing the motional mass and spring constant of the sensing mode, and therefore combines the sensitivity of a nanowire detector with the rigidity of a cantilever beam.  We experimentally demonstrate the performance of the nanoladder cantilever by achieving a force noise of $158\unit{zN/\rtHz}$ at $108\unit{mK}$. 


We implemented the nanoladder design using diamond and silicon as resonator materials.  Diamond devices were patterned by electron beam lithography onto a polished 20-$\um$-thick single crystal membrane and released by inductively coupled plasma (ICP) etching from the backside\cite{Tao2013,Tao2014}.  Silicon devices used a low-conductivity SOI wafer and a vapor hydrofluoric acid (HF) etching process for device formation (see SI).  In order to protect the fragile nanoladder structure during the final release, most of the cantilevers were tethered to the opposite support substrate (see Figure \ref{fig:Fig1}).  The tether was cut post-release by a focused helium ion beam (diamond devices) or mechanically with a manual micromanipulator stage (silicon devices).  Optical and scanning electron micrographs of a finished diamond device are shown in Figure \ref{fig:Fig1}(a); further images are provided in the SI. 
%


We characterized the nanoladder cantilevers in a custom-built force microscope operating inside the vacuum chamber of a dilution refrigerator (Leiden CF450) with a minimum stage temperature of approximately $80\unit{mK}$ \cite{Tao2014}.  The microscope stage was suspended on springs to eliminate environmental vibrations.
The subnanometer vibrations of the cantilevers were measured by a fiber-optic interferometer operating at a wavelength of 1550 nm (Fig. \ref{fig:Fig1}(b)) \cite{Rugar1989}.  Only a few percent of the light incident at the cantilever were typically reflected back into the fiber because of the small size of the mirror paddle, and because the interferometer wavelength was not well matched to the cantilever thickness \cite{Mamin2001,Mulhern1991}.


%
\begin{figure}
	\includegraphics[width=\columnwidth]{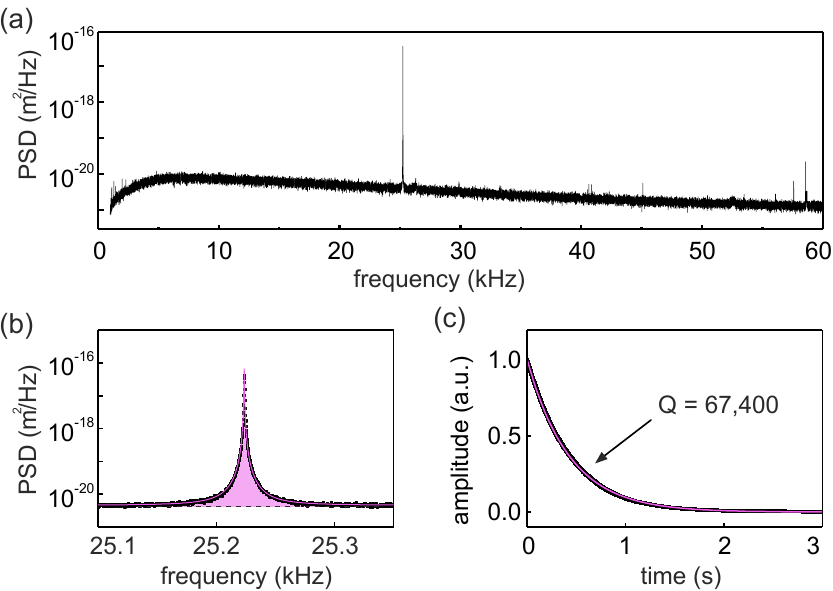}
	\caption{Characterization of a diamond nanoladder cantilever.
	(a) Power spectral density (PSD) of the cantilever displacement measured at room temperature.
	(b) PSD of the fundamental mode at $\fc = 25.22\unit{kHz}$.  The shaded area corresponds to $\xrms^2$.  The baseline reflects the displacement noise of the interferometer.
	(c) Ring-down experiment used to determine the $Q$ factor at room temperature (average over $100$ ring-downs).
	}
	\label{fig:Fig2}
\end{figure}
In a first step, we measured the vibrational noise spectrum at room temperature (Figure \ref{fig:Fig2}(a)).  The spectrum was dominated by a sharp signal peak at the frequency $\fc$ of the fundamental mechanical mode, reflecting the thermomechanical motion of the cantilever.  The higher order modes appeared at much larger frequencies (see SI), as expected from the high transverse and longitudinal stiffness of the nanoladder.  To determine the spring constant $k$ and effective mass $m$ of the fundamental mode, we computed the mean square displacement $\xrms^2$ by integrating the signal peak of the power spectral density (PSD) after subtracting the background (Fig. \ref{fig:Fig2}(b)).  In thermal equilibrium, the mean square displacement is linked to the spring constant by the equipartition theorem, $\tfrac12 k\xrms^2 = \tfrac12 \kT$, where $T$ is the noise temperature of the resonant mode \cite{Mamin2001}.  By performing this measurement at room temperature ($T=295\unit{K}$) we could calibrate both the spring constant, $k=\kT/\xrms^2$, and the effective mass, $m = k/(2\pi\fc)^2$.  The quality factor $Q$ was determined via the decay time constant $\tau=Q/(\pi\fc)$ of a separate ringdown measurement (Fig. \ref{fig:Fig2}(c)).

\begin{table*}[t!]
\centering
\begin{tabular}{lcccccl}
\hline\hline
\rowcolor{white}[1.2\tabcolsep] Device & $\fc$ (kHz) & $m$ (pg) &  $\kc$ ($\uN/m$) & $Q^a$ & $\gamma$ (pg/s) & Reference \\
\hline\hline
\rowcolor{sorange}[1.2\tabcolsep] Nanoladder (diamond) & $25.22$ & $4.1 \pm 0.6$ &  $110 \pm 10$   & $162,000$ & $3.7$ & This work \\
\rowcolor{white}[1.2\tabcolsep] Rectangular beam (diamond) & $32.14$ & $1,600$ & $67,000$ &  $6,000,000$ & $55$ & Tao \etal \cite{Tao2014}    \\
\hline\hline
\rowcolor{sblue}[1.2\tabcolsep] Nanoladder (silicon) &  $5.52$ & $5.5 \pm 1.3$ &    $6.5 \pm 1.6$ &  $45,000$ & $4.2$ & This work \\
\rowcolor{white}[1.2\tabcolsep] Rectangular beam (silicon) &  $4.98$ &   $270$ &    $260$ &    $150,000$ & $55$ & Mamin \etal \cite{Mamin2001} \\
\hline\hline
\end{tabular}
\caption{Mechanical properties of nanoladder cantilevers and corresponding rectangular beam cantilevers.  Data for the diamond cantilever are for the device shown in Fig. \ref{fig:Fig1}. A total of 5 diamond and 2 silicon devices were investigated (see SI).
$^a$ Reported $Q$ factors represent the highest measured values at millikelvin temperatures.}
\label{fig:table1}
\end{table*}

Table \ref{fig:table1} compares the mechanical properties of a diamond and a silicon nanoladder to state-of-the-art ultrasensitive cantilever beams (Refs. \onlinecite{Mamin2001}, \onlinecite{Tao2015}).  We find that the nanoladder devices have $50-400$ times lower mass and are $40-600$ times softer than their rectangular beam counterparts.  Ideally, the nanoladder design should therefore allow for a reduction of the mechanical dissipation $\gamma = \sqrt{k m}/Q$ by roughly two orders of magnitude.  However, since the nanoladders also exhibit a significantly lower $Q$ factor than the corresponding rectangular beam devices, the reduction in mechanical dissipation is effectively about a factor of 15.  The reduction in the $Q$ factor is probably related to the enhanced surface-to-volume ratio and to sidewall roughness, and will be the subject of a future study.

\begin{figure}
	\includegraphics[width=\columnwidth]{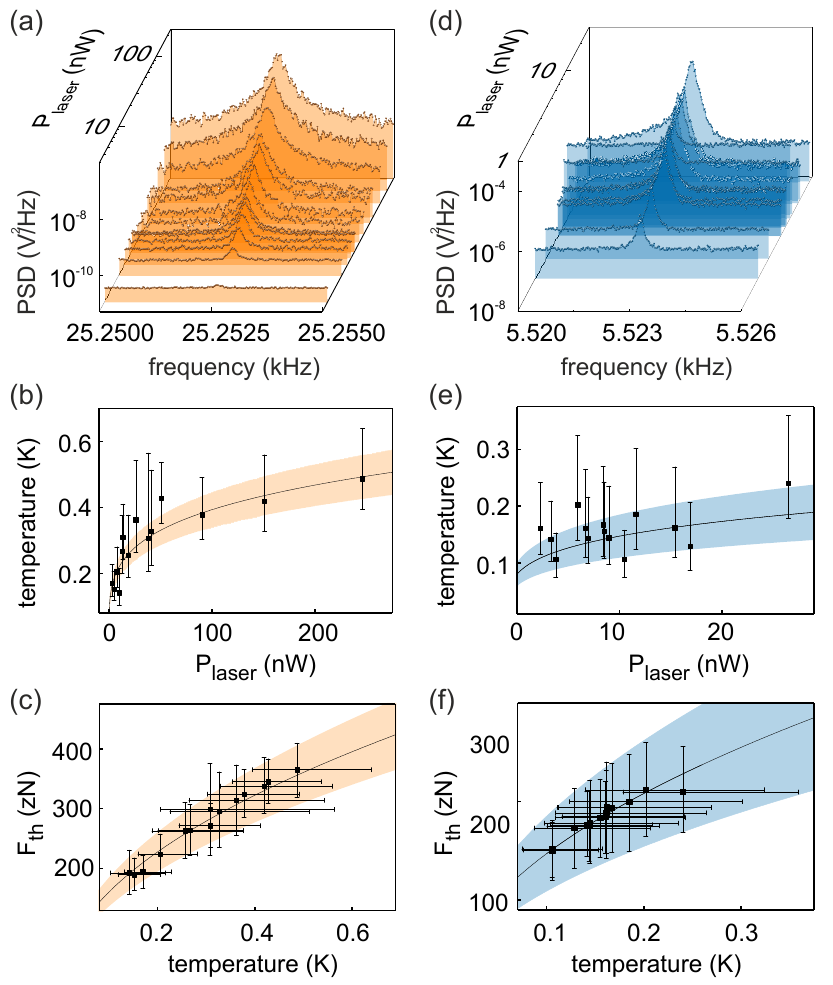}
	\caption{Characterization of nanoladder cantilevers at cryogenic temperatures.
	(a) Diamond device: thermal displacement PSD as a function of interferometer laser power$^a$. 
	(b) Mode temperature as a function of interferometer laser power.
	(c) Thermal force noise (in a one-Hz bandwidth) as a function of mode temperature. 
	(d-f) Corresponding data for the silicon device.
	Error bars reflect the fit error and shaded areas indicates calibration uncertainty in the resonator mass (see SI).
	$^a$ Displacement PSD is given in units of V$^2$/Hz because the displacement sensitivity varied between measurements.
	}
	\label{fig:Fig3}
\end{figure}
We now turn to the main result of this work, a measurement of the force noise of nanoladder cantilevers at cryogenic temperatures.  For this purpose, the devices were cooled down to the base temperature of the dilution refrigerator.  Below $\sim 1\unit{K}$, we observed that the mode temperature $T$ of the cantilevers was higher than the stage temperature $\Tstage$ due to laser absorption at the paddle.
To study the cantilever noise as a function of temperature, we therefore kept the stage temperature constant ($\Tstage \sim 80\unit{mK}$) and varied the power $\Plaser$ of the laser light injected into the interferometer arm (see Figure \ref{fig:Fig1}(b)).

Figure \ref{fig:Fig3}(a) shows displacement spectra for the diamond cantilever as a function of $\Plaser$.  For each spectrum, we calculated the mode temperature $T = m\xrms^2 (2\pi\fc)^2/\kB$ by integrating the spectrum (after baseline subtraction) and using the known mass from the room temperature calibration measurement.  The mode temperature roughly followed $T = (\epsilon\Plaser+ \Tstage^4)^{1/4}$ as the laser power was reduced from $\Plaser = 250\unit{nW}$ to $1\unit{nW}$, as expected from the thermal conductivity of diamond at low temperatures \cite{Tao2014} [lines in Figs. 3(b,e)].
The quality factor concurrently increased from $Q=$131,000 to 162,000 (see SI).
At the lowest temperature of $T\sim 140\unit{mK}$, the force noise reached a value of $\Fth = 190 \substack{+42 \\ -33} \unit{zN}$ in bandwidth $B=1\unit{Hz}$.
Note that because the experimenter also sees detector noise in addition to the force noise, the value for the force sensitivity is slightly higher, $\Fmin = 340\substack{+75 \\ -59}\unit{zN}$ (see SI).  The dominant error in the force resolution is the calibration error of the cantilever displacement, which leads to an uncertainty in the estimate of the resonator mass (see SI and shaded areas in Fig. \ref{fig:Fig3}).

Figures \ref{fig:Fig3}(d-f) show the corresponding data for the silicon devices.  Here, the minimum force noise is $\Fth = 158 \substack{+62 \\ -42}  \unit{zN}$ and the force sensitivity is $\Fmin = 188 \substack{+70 \\ -48} \unit{zN}$ at a mode temperature of $T\sim 110\unit{mK}$, almost identical to the diamond cantilever.  These results demonstrate that the nanoladder design provides a generic improvement of the force noise independent of the resonator material.

\begin{table*}[t!]
\centering
\begin{tabular}{l|l|ccc|l}
\hline\hline
\rowcolor{white}[1.2\tabcolsep]  &  & \multicolumn{3}{c}{$\Fth$ (aN)} & \\
\rowcolor{white}[1.2\tabcolsep] Geometry & Material & $300\unit{K}$ & $3\unit{K}$$^a$ & $100\unit{mK}$ & Reference \\
\hline\hline
\rowcolor{sorange}[1.2\tabcolsep] Nanoladder & diamond & 13 & 1.1 & 0.19 & this work \\
\rowcolor{sblue}[1.2\tabcolsep] Nanoladder & silicon & 15 & 1.2 & 0.16 & this work \\
\rowcolor{white}[1.2\tabcolsep] Singly-clamped beam & diamond & 115 & 6.0 & 0.54 & Tao \etal \cite{Tao2014} \\
\rowcolor{white}[1.2\tabcolsep] Singly-clamped beam & silicon & --- & 4.7 & 0.82 & Mamin \etal \cite{Mamin2001} \\
\hline
\rowcolor{white}[1.2\tabcolsep] Nanowire & silicon & --- & 1.5 & --- & Nichol \etal \cite{Nichol2012} \\
\rowcolor{white}[1.2\tabcolsep] Nanowire & GaAs/AlGaAs & --- & 5 & --- & Rossi \etal \cite{Rossi2017} \\
\rowcolor{white}[1.2\tabcolsep] Membrane & Si$_3$N$_4$ & 19.5 & --- & --- & Reinhard \etal \cite{Reinhard2016} \\
\rowcolor{white}[1.2\tabcolsep] Membrane & Si$_3$N$_4$ & (10)$^b$ & --- & --- & Norte \etal \cite{Norte2016} \\
\hline\hline
\end{tabular}
\caption{Reported sensitivities in bandwidth $B=1\unit{Hz}$ for ultrasensitive cantilevers.  Prospective nanowire and membrane devices are included for reference.
$^a$ temperatures vary between 1-8 K. $^b$ calculated (not measured).
}
  \label{fig:table2}
\end{table*}

Table \ref{fig:table2} compares the force sensitivities of the nanoladder cantilevers to the cantilever beams of Ref. \onlinecite{Mamin2001} and \onlinecite{Tao2014}.  Clearly, the nanoladder design provides a significant improvement compared to standard cantilever beams.
In fact, to the best of our knowledge, our experiments observed the lowest force noise that has been reported for a cantilever device.
The nanoladder devices also equal or surpass the sensitivities offered by other prospective geometries considered for force microscopy, such as singly-clamped nanowires \cite{Nichol2012,Rossi2017} or two-dimensional membranes \cite{Reinhard2016,Norte2016}, regardless of the experimental temperature.  The sensitivity of the nanoladder devices could be further improved by lowering the mode temperature, which requires a suppression of the laser heating.  This can be achieved by relatively simple means.  A higher reflectivity of the cantilever mirror paddle, for example, will at the same time increase the signal amplitude and decrease the effect of absorptive heating, leading to lower mode temperatures.  For this purpose, future generations of nanoladder cantilevers might be equipped with a highly reflective mirror on the paddle \cite{Kemiktarak2012}.  It may also be possible to further improve the sensitivity by reducing the intrinsic dissipation of the devices, for example by applying a chemical surface treatment \cite{Tao2015}, reducing sidewall roughness, or by adopting phonon bandgap engineering ideas from optomechanical membranes \cite{Yu2014,Tsaturyan2017}.

Looking forward, the nanoladder cantilever design can enable a significant step ahead in the mechanical detection of weak magnetic signals, such as those produced by nuclear or electronic spins in the context of magnetic resonance force microscopy \cite{Poggio2010}.  For instance, the force generated on a single proton magnetic moment ($\mu = 1.4\ee{-26}\unit{A m}$) by the field gradient of a strong nearby nanomagnet ($G = 2.8\ee{7}\unit{T/m}$, Ref. \onlinecite{Tao2016}) is $F=\mu G = 400\unit{zN}$, which is more than two times larger than the sensitivity limit (in a one Hz bandwidth) achieved in this work.  It is therefore conceivable that nanoladder cantilevers will pave the way towards single nuclear spin detection, which is an important prerequisite for the three-dimensional magnetic resonance imaging of individual molecules with atomic resolution \cite{Sidles1991,Degen2009}.  Diamond nanoladders with low dissipation are also of interest to on-going efforts in optomechanics to exploit the excellent photonic properties of the material and its unique intrinsic lattice defects \cite{Rabl2009}.

\begin{acknowledgements}
The authors thank Christoph Keck for experimental assistance.
This work has been supported by the ERC through Starting Grant 309301, by the European Commission through the FP7-611143 DIADEMS programme, by the Swiss NSF through the NCCR QSIT, and by the ETH Research Grant ETH-03 16-1.
Y.P. and Y.T. acknowledge funding from a Rowland Fellowship.
I.S. is thankful to Empa for financial support and to Swiss National Science Foundation for support in equipment procurement (REquip 206021\_133823).
\end{acknowledgements}

%

\clearpage
\onecolumngrid

\large
\begin{center}
\textbf{\large Supplemental Materials: Nanoladder cantilvers made from diamond and silicon}\\
\normalsize
M. H\'{e}ritier$^{1}$, A. Eichler$^{1}$, Y. Pan$^{2}$, U. Grob$^{1}$, I. Shorubalko$^{3}$, M. D. Krass$^{1}$, Y. Tao$^{2}$ and C. L. Degen$^{1}$
\vspace{5 mm}

{\small\noindent
$^1$Department of Physics, ETH Zurich, Switzerland\\
   $^2$Rowland Institute at Harvard, Cambridge MA, USA\\
	$^3$Transport at Nanoscale Interfaces, Empa, Swiss Federal Laboratories for Materials Science and Technology, Duebendorf, Switzerland\\
}
\end{center}
\small
\makeatletter
\renewcommand{\theequation}{S\arabic{equation}}
\renewcommand{\thefigure}{S\arabic{figure}}
\renewcommand{\thetable}{S\arabic{table}}
\renewcommand{\bibnumfmt}[1]{[S#1]}
\renewcommand{\citenumfont}[1]{S#1}
\setcounter{equation}{0}
\setcounter{figure}{0}
\setcounter{table}{0}
\renewcommand{\figurename}{\textbf{Supplementary Figure}}
\renewcommand{\tablename}{\textbf{Supplementary Table}}

\section{Device fabrication}
\subsection{Fabrication of single-crystal diamond nanoladders}
\begin{figure}[h]
\centering
	\includegraphics[width=10cm]{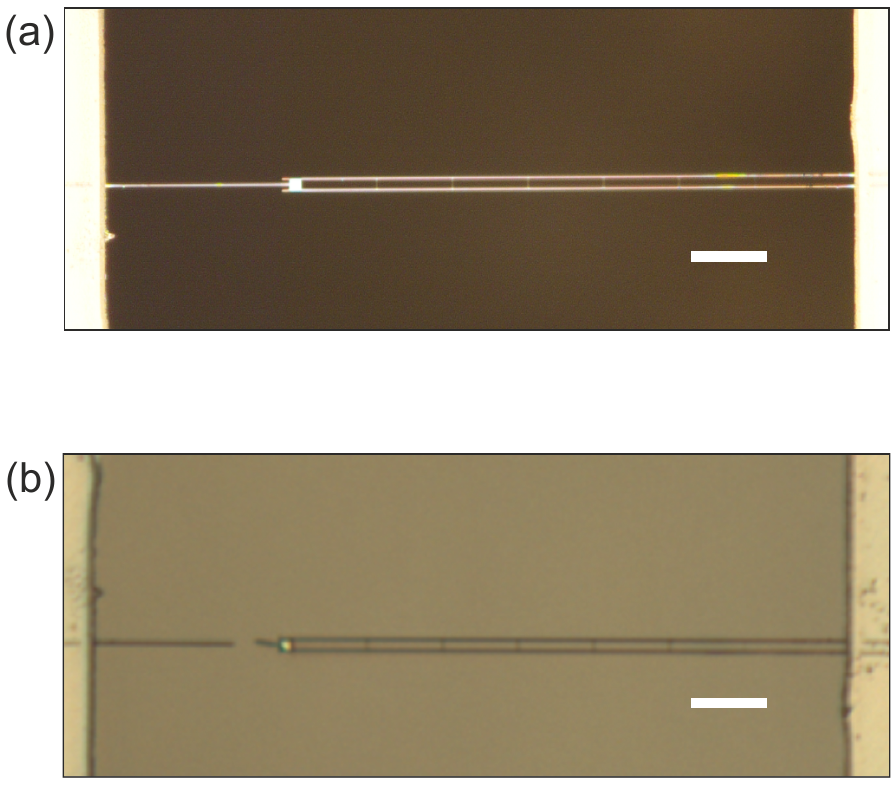}
	\caption{Focused ion beam cutting of diamond nanoladders. (a) Optical micrograph of a $150\unit{\um}$ diamond nanoladder cantilever after the final release. In order to protect the fragile structure, it was tethered to the opposite support substrate. (b) Optical micrograph of the same device after cutting by a focused helium ion beam which offers a high precision while inflicting minimal material damage. Scale bars are $20\unit{\um}$.}
	\label{fgr:imagesdevices}
\end{figure}

Diamond devices were patterned by electron beam lithography onto an electronic-grade single crystal with a (100) surface orientation that was polished down to 20 $\um$ thickness.  The pattern was transferred to the diamond and released by inductively coupled plasma (ICP) etching. Detailed instructions can be found in Ref. S1 and S2 with the important difference that the nanoladders were tethered to the support substrate (Fig. \ref{fgr:imagesdevices} (a)), in order to protect the fragile structure of the diamond nanoladders during the final release step. To cut the tether after the release, we used a focused helium ion beam (He-FIB) (Fig. \ref{fgr:imagesdevices} (b)). Zeiss Orion Plus He-FIB [S3] equipped with a Raith Elphy MultiBeam pattern generator were used for presented study. He-FIB was operated at $30\unit{kV}$ acceleration voltage and $\sim 5\times 10^{-5}\unit{Pa}$ chamber pressure. He-ion beam current selected for this work was $5-10\unit{pA}$. It was defined by the following hardware parameters: $10$ micrometer aperture, $6.7 \times 10^{-4}\unit{Pa}$ helium pressure in the gun chamber, and “spot” parameter between $2.5$ and $3$. The diamond cantilevers were cut by irradiating a box $500\times 100 \unit{nm}$ with a dose $\sim 1\unit{C/cm^2}$. Such dose is expected to make a $200-400\unit{nm}$ deep cut [S4]. 
\subsection{Fabrication of single-crystal silicon nanoladders}
\begin{figure}[h]
\centering
	\includegraphics[width=10cm]{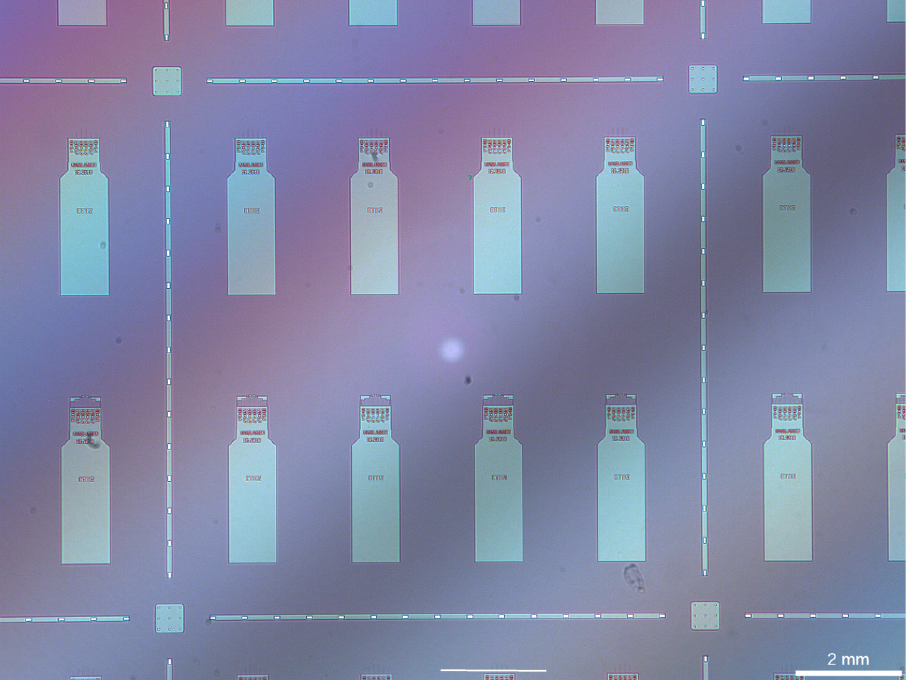}
	\caption{Wafer-scale fabrication of silicon nanoladders. Optical micrograph of chips carrying 6 nanoladders of different lengths after electron beam lithography and inductively-coupled plasma etching step. The chip dimensions can be easily changed during fabrication to allow mounting them in a variety of systems. Scale bar: $2\unit{mm}$}
	\label{fgr:fab}
\end{figure}
Silicon devices were batch-fabricated using single crystalline silicon-on-insulator (SOI) wafers (Soitec) with $1500\pm30\unit{nm}$ device layer, $1000\pm10\unit{nm}$ buried oxide layer, and $725\pm15\unit{\um}$  handle layer. The device silicon layer was p-type, (100)-oriented, and had a resistivity of $18\pm4 \unit{Ohm\cdot cm}$.
\begin{figure}[h]
\centering
	\includegraphics[width=10cm]{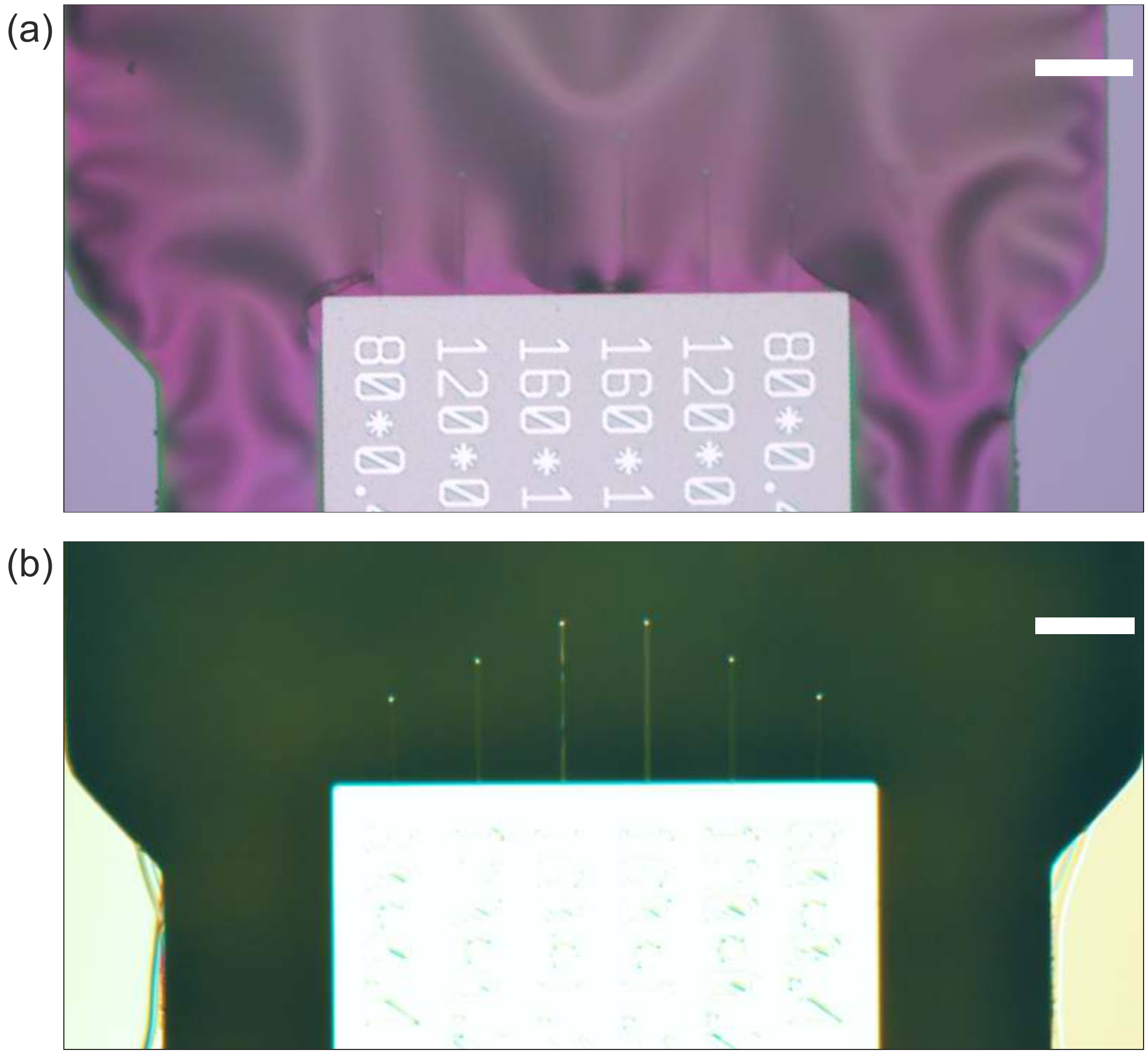}
	\caption{Final fabrication steps for silicon nanoladders. (a) Chip before the final vapor HF etch to remove the buried oxide layer and thus releasing the nanoladders. Scale bar: $100\unit{\um}$ (b) Ready-to-use chip after the final fabrication step. The chips can be easily picked out with a pair of tweezers and mounted into our custom-built AFM stage. Scale bar: $100\unit{\um}$}
	\label{fgr:fab2}
\end{figure}
A PECVD silicon nitride (SiNx) layer with $260\pm10$ nm thickness was deposited on the device layer of a $4$-inch SOI wafer and patterned by e-beam lithography (EBL). The device silicon layer was then wet-etched in tetramethyl ammonium hydroxide (TMAH) solution ($25\%$, heated to $60\,^\circ$C) with the patterned SiNx layer as a hard mask. The thickness of the device layer was monitored at different locations using profilometry in regular intervals during the silicon thinning process. This step determined the thickness of the nanoladders.

A layer of hydrogen silsesquioxane (HSQ), negative e-beam resist solution ($6\%$) with $170$ nm thickness was spin-coated ($500$ rpm, $5$ s; $2000$ rpm, $60$ s) onto the SOI wafer with the thinned device layer. E-beam lithography was used to write the nanoladder features. After exposure, the sample was developed in a 351B:H2O solution (volume ratio $1:3$) for $5$ minutes. The resulting pattern served as a mask during the subsequent inductively coupled plasma (ICP) etching step, in which HBr plasma in the ICP was used to transfer the HSQ pattern into the underlying device layer. This is possible because HSQ is five times more etch-resistant than silicon. The residual HSQ mask was then removed by a short dipping ($15$ s) in buffered hydrofluoric acid (BHF). At this point, the silicon nanoladders remain anchored to the buried oxide layer (see Fig. \ref{fgr:fab} (a)).

The wafer was ready for backside etching to suspend the silicon nanoladders. To protect the patterned structures, AZ4562 resist was spin-coated at $4000$ rpm onto the front side of the wafer. Thicker AZ4562 resist was spin-coated at $1500$ rpm onto the backside to provide sufficient etch-resistance during DRIE. Following backside alignment and photolithography, the wafer was glued using wax to a dummy silicon carrier wafer via the front side. DRIE of the handle-layer silicon was performed to etch through the silicon down to the buried oxide stop layer.  

To release the nanoladders, the wafer was soaked in warm NMP at $50\,^\circ$C.  The wafer was cleaned by rinsing, sequentially, in the following solvents: acetone, IPA, DI water, piranha solution, DI water, and IPA.  The wafer was allowed to dry in air after taking it out from the final IPA rinse.  To minimize contamination by drying residues, care was taken to keep the wafer tilted on a piece of cleanroom wipe to allow as much liquid to flow away as possible.  Nitrogen flow was not used to avoid damaging the delicate buried oxide membranes. The wafer was etched in vapor HF for $10$ minutes to remove the buried oxide layer, thus releasing the silicon nanoladders. Figures \ref{fgr:fab2} (b) and (c) show optical micrographs of a wafer before and after this final step.

\begin{figure}[h]
\centering
	\includegraphics[width=10cm]{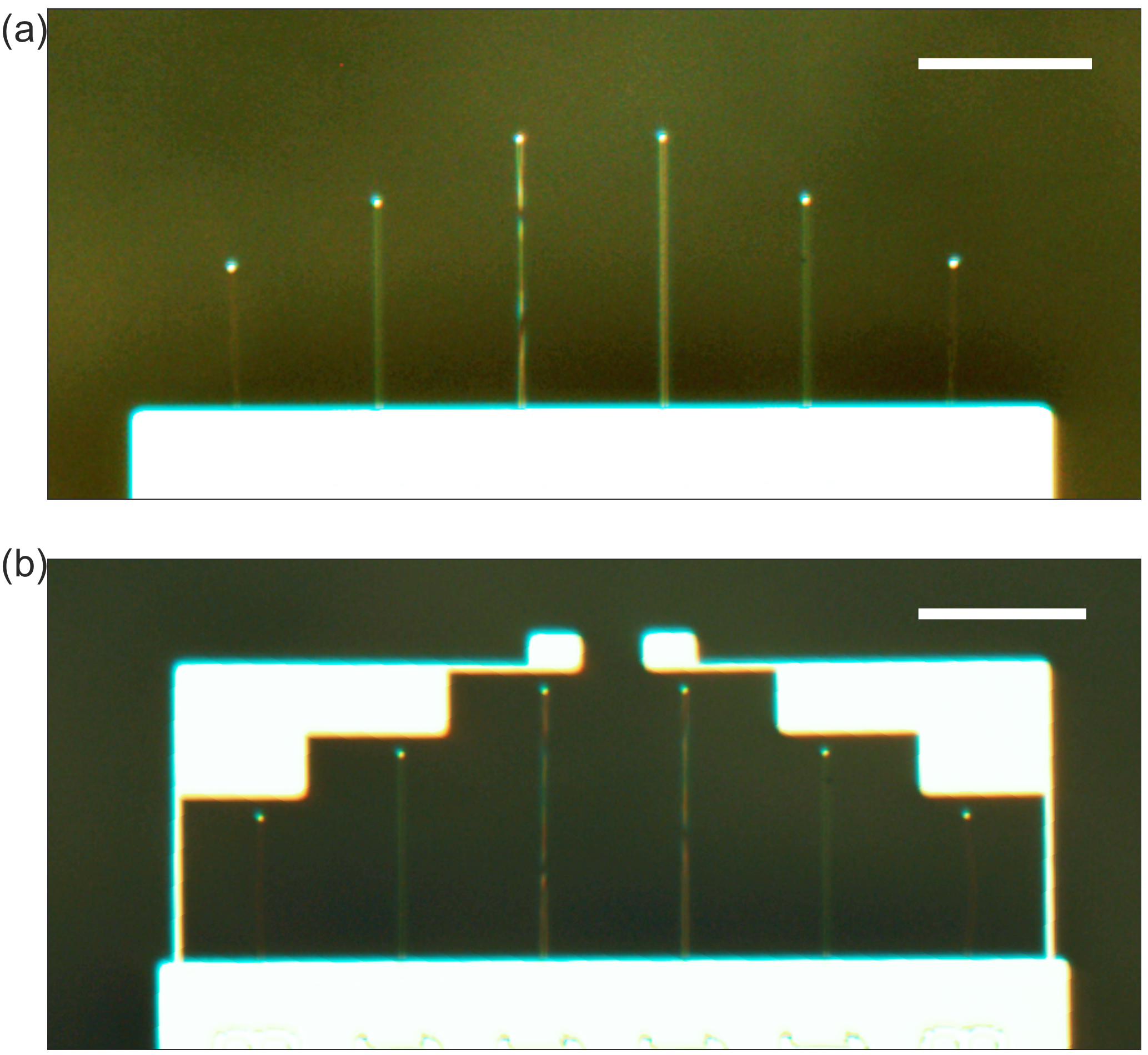}
	\caption{Tethering of devices. (a) Chip with finished silicon devices fabricated without tether. (b) Same devices but fabricated with help of a tether. Scale bars are $100\unit{\um}$.}
	\label{fgr:fab3}
\end{figure}
Taking advantage of the wafer-scale production, we designed some silicon nanoladders with a tether and some without (see Fig. \ref{fgr:fab3}). While the tethered devices could have been cut by a focused ion beam, a hydraulic micromanipulator system (Narishige Three-axis Hanging Joystick Oil Hydraulic Micromanipulator, model MMO-202ND) operated under an optical microscope yielded the same result.
\begin{figure}[h]
\centering
	\includegraphics[width=10cm]{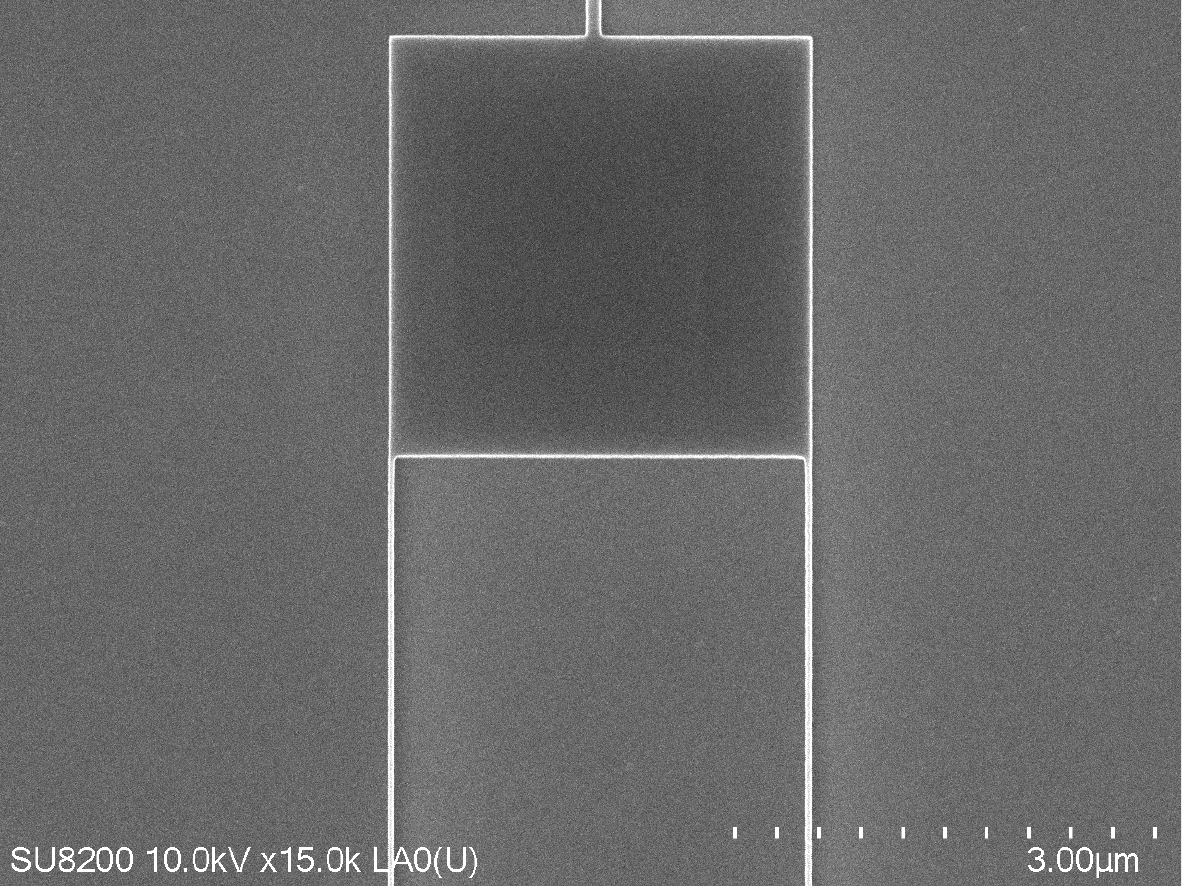}
	\caption{End mirror of the nanoladders. SEM micrograph of the paddle used for reflecting the readout laser beam from the cantilever. With a paddle area of $3\times 3\unit{\um^2}$, the aspect ratio between the cantilever and the arms width is extreme. In this case, the arms width and thickness lie in the range of $50-60 \unit{nm}$, smaller than for the devices presented in the main manuscript which have a width and thickness in the range of $100-300 \unit{nm}$.}
	\label{fgr:imagesdevices2}
\end{figure}

\clearpage
\section{Characterization of further devices}
\begin{figure}[h]
\centering
	\includegraphics[width=15cm]{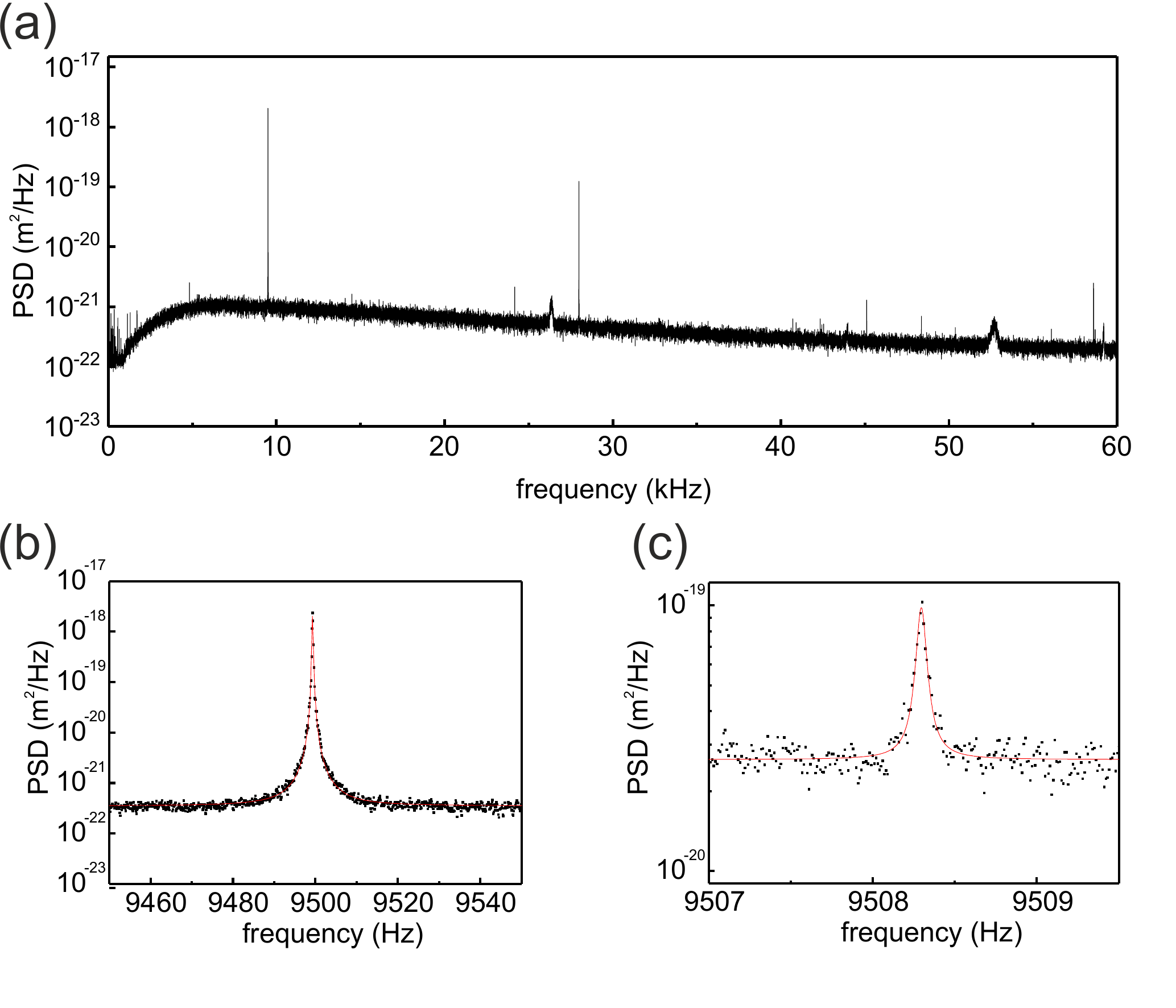}
	\caption{Characterization of a $300\unit{\um}$-long diamond nanoladder cantilever (cantilever $\# 5$ in table \ref{tabledevices}). (a) Power spectral density (PSD) of the cantilever displacement measured at room temperature. (b) and (c) PSD of the fundamental mode ($f_0=9.499\,$kHz, $f_0=9.508\,$kHz) at room and cryogenic temperatures, respectively.}
	\label{fgr:qvst}
\end{figure}

We have investigated a total of 5 diamond and 2 silicon devices whose characteristics are shown in table \ref{tabledevices}. In particular, a $300 \unit{\um}$-long diamond cantilever has been characterized in the same manner as the devices of the main manuscript. It shows a similar behavior and agrees as well with our simulations. Its characteristics are summarized in table \ref{table2}. The resonance frequency of the fundamental mode, $f_0 = 9,496.1\,$Hz, is measured from the thermomechanical displacement power spectral density shown in fig \ref{fgr:qvst}a.
\begin{table}[h]
\centering
\caption{Characteristics of different devices at room temperature.}
\label{tabledevices}
\begin{tabular}{ccccc}
\hline
 cantilever $\#$ & material & Length  ($\unit{\um}$)            & $f_c$ ($\unit{kHz}$) & $Q$  \\ \hline
\rowcolor{sorange}[1.2\tabcolsep] $1$ & diamond & $100$ & $25.22$ & $60,000$ \\ \hline
$2$ & diamond & $150$ & $13.914$ & $37,000$ \\ \hline
$3$ & diamond & $250$ & $4.640$ & $48,000$ \\ \hline
$4$ & diamond & $300$ & $4.180$ & $40,000$ \\ \hline
\rowcolor{sgreen}[1.2\tabcolsep] $5$ & diamond & $300$ & $9.500$ & $62,000$ \\ \hline \hline
$6$ & silicon & $160$ & $4.282$ & $14,000$ \\ \hline
\rowcolor{sblue}[1.2\tabcolsep] $7$ & silicon & $160$ & $5.523$ & $12,000$ \\ \hline
\end{tabular}
\end{table}

\begin{table}[h]
\centering
\caption{Key characteristics of a $300\,\rm\mu$m long single crystal diamond nanoladder cantilever (cantilever $\# 5$ in table \ref{tabledevices}). $^a$ Reported Q factor represent the highest measured value at milikelvin temperatures.}
\label{table2}
\begin{tabular}{ccccc|ccc}
\hline\hline
& & & & & \multicolumn{3}{c}{$\Fth$ (aN/$\rtHz$)} \\
$\fc$ (kHz) & $m$ (pg) &  $\kc$ ($\uN/m$) & $Q^a$ & $\gamma$ (pg/s) & $300\unit{K}$ & $4\unit{K}$ & $100\unit{mK}$ \\
\hline\hline
 \rowcolor{sgreen}[1.2\tabcolsep] 9.5 & 19.3 & 69 &  161,000 & 7.2 & 18 & 1.6 & 0.27 \\
\hline\hline
\end{tabular}
\end{table}

\clearpage
\section{Finite element modeling}
\subsection{Mode shape simulation}
\begin{figure}[h]
\centering
	\includegraphics[width=10cm]{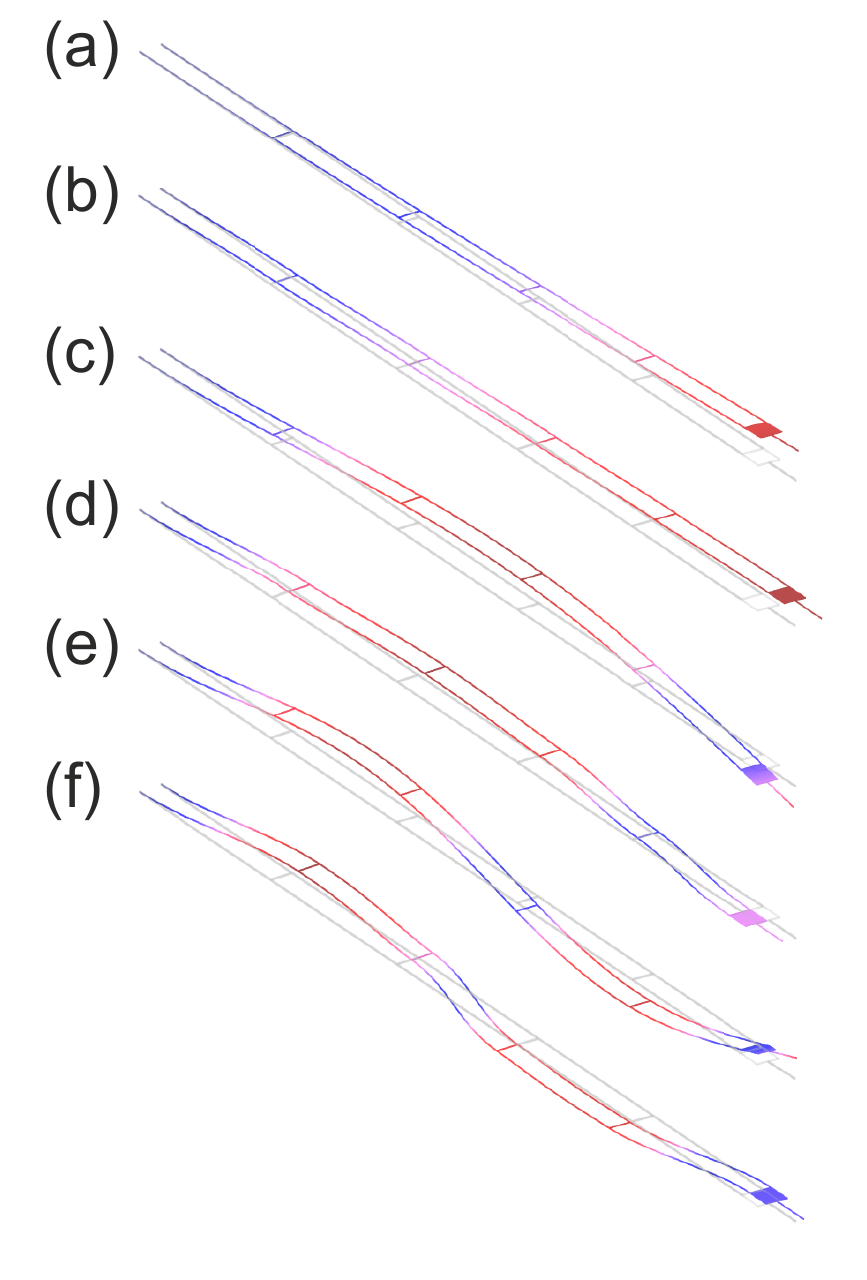}
	\caption{COMSOL mode shape simulation of the fundamental and higher modes. (a) Fundamental mode, $f_{c}^{(a)} = 25.4\unit{kHz}$, (b) $f_{c}^{(b)} = 142\unit{kHz}$, (c) $f_{c}^{(c)} = 204\unit{kHz}$, (d) $f_{c}^{(d)} = 488\unit{kHz}$, (e) $f_{c}^{(e)} = 619\unit{kHz}$, (f) $f_{c}^{(f)} = 905\unit{kHz}$. (a), (c), and (f) correspond to in-plane motion whereas (b), (d), and (e) include only lateral motion.}
	\label{fgr:comsolsimul}
\end{figure}

We performed finite element simulations to determine the mode shape of the nanoladders and their resonance frequencies. Figure \ref{fgr:comsolsimul} shows the results where (a) corresponds to the fundamental resonance frequency, and (b)-(f) show increasingly higher modes. (a), (c), and (f) correspond to in-plane motion whereas (b), (d), and (e) include only lateral motion. The fundamental mode has approximately the same shape as the one of a plain beam cantilever.
\subsection{Estimations of resonance frequency and spring constant}
We compare our results to theoretical simulations by two different approaches. In a first approximation, the rungs and the paddle are neglected and the nanoladder is considered as two single beams coupled in parallel. The effective spring constant $k$ writes then $k=2 k_{\rm beam}$ where $k_{\rm beam}$ is the spring constant of a single beam. It is given by 
\begin{equation}
k_{\rm beam} = W \left( \frac{T}{L} \right)^3 \frac{E}{4}
\end{equation} with $W$ being the width, $T$ the thickness, $L$ the length and $E$ the Young's modulus of the beam. The fundamental mode frequency reads 
\begin{equation}
f_0 = 0.162  \frac{T}{L^2} \sqrt{ \frac{E}{\rho} }
\end{equation} where $\rho$ is the Poisson's ratio of the material. Fixing the length to $100\,\rm\mu$m or $300\,\rm\mu$m, respectively, we computed lower and upper boundaries for frequency and spring constant by setting the width and the thickness to $100\,$nm and $300\,$nm, respectively.

The second method is a finite element simulation performed for the same parameters through COMSOL. The simulated results are shown in table \ref{tablesimul} and are compared to the measured values.
\begin{table}[h]
\centering
\caption{Simulated cantilever characteristics compared to the measured values. \\ $^a$Device $\#$ in table \ref{tabledevices})}
\label{tablesimul}
\begin{tabular}{c|ccc|ccc}
\hline\hline
Device$^a$ & \multicolumn{3}{c}{$\fc$ (kHz)} & \multicolumn{3}{c}{$\kc$ ($\uN/m$)} \\
 & analytical & COMSOL & measured & analytical & COMSOL & measured \\
\hline\hline
 \rowcolor{sorange}[1.2\tabcolsep] $\# 1$ & 30.1-90.4 & 16.9-68.0 &  25.2 & 61-4940 & 85-8918 & 106 \\
	\rowcolor{sgreen}[1.2\tabcolsep] $\# 5$  & 3.35-10 & 2.45-22.6 &  9.50 & 2.26-183 & 3.90-404 & 69 \\
		\rowcolor{sblue}[1.2\tabcolsep]  $\# 7$  & 5.06-15.2 & 3.30-12.5 & 5.46 & 10.2-710 & 2.88-297 & 6.5 \\
\hline\hline
\end{tabular}
\end{table}

\clearpage
\section{Quality factor study}
\begin{figure}[h]
\centering
	\includegraphics[width=10cm]{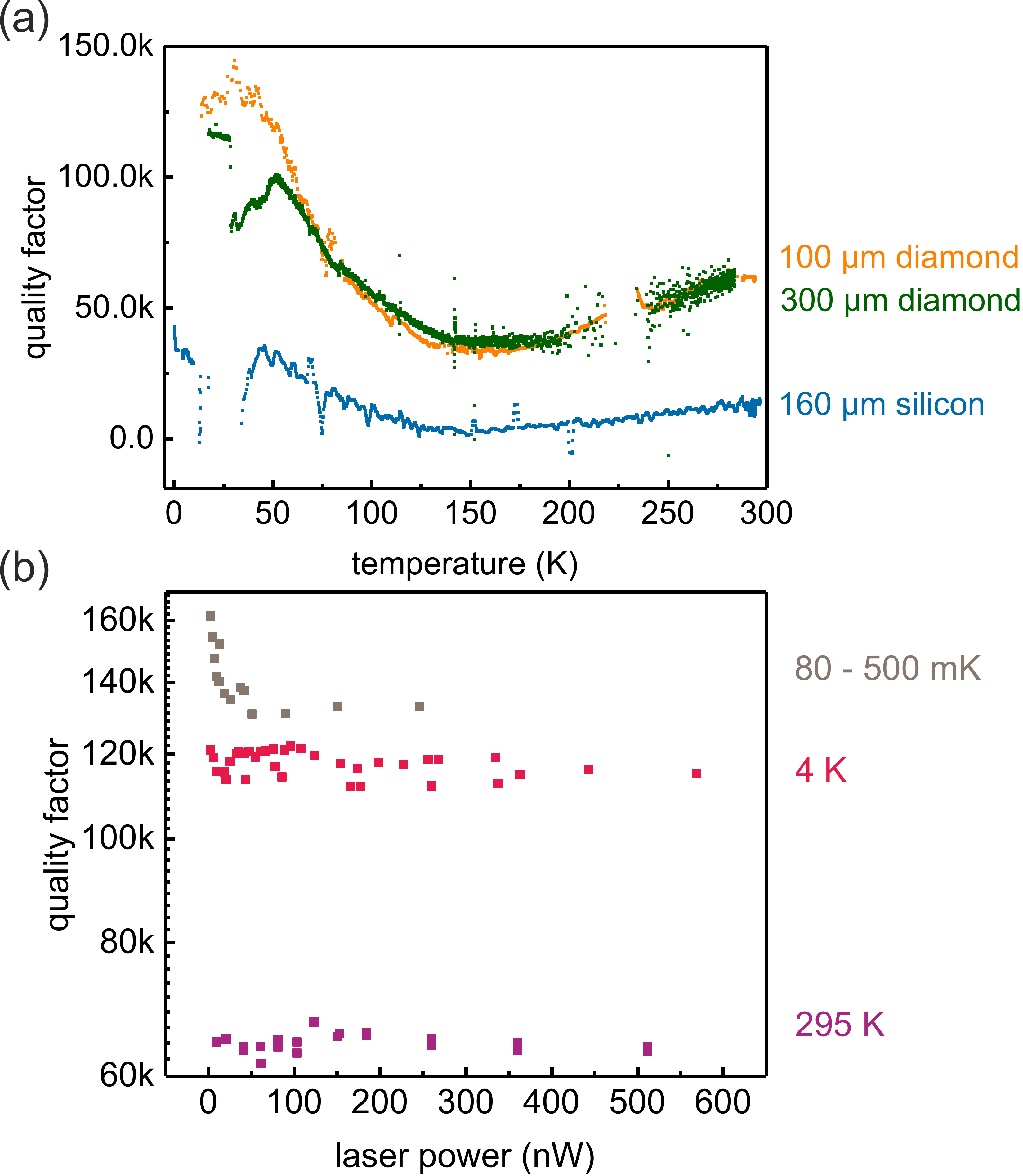}
	\caption{Quality factor measurements. (a) $Q$ between $4$ and $300$ K. Orange corresponds to device $\# 1$, green to $\# 5$ and blue to $\# 6.$ (b) $Q$ as function of the laser power at different temperature ranges. The purple squares correspond to values at room temperature, the red ones were taken at $4$ K, and the light gray denote milikelvin temperature measurements. While no change is observed for $300$ and $4$ $\unit{K}$, $Q$ increases with decreasing laser power in the milikelvin range.}
	\label{fgr:qvstemp}
\end{figure}

 We have studied the quality factor dependance on temperature for our different devices as shown in fig \ref{fgr:qvstemp} (a). It increases from room temperature towards cryogenic temperatures for all 3 devices.  The dip around $150$\,K has been previously observed in single-crystal diamond (electronic and optical grade), polycrystalline diamond and single-crystal silicon. The feature has been ascribed to dissipation caused by surface adsorbates or a surface passivation layer~[S2, S5]. Around $230$\,K, an interferometer malfunction prevented a precise measurement and we therefore removed the data in the relevant temperature range.

As shown in the main manuscript, the mode temperature of the cantilever depends on the incident laser power in the milikelvin range. During this measurement, we have monitored the quality factor as well as seen in fig \ref{fgr:qvstemp} (b). The quality factor increases for decreasing mode temperatures. No change of the quality factor is observed at room temperature and at $4$ K. This corresponds to the expectation that the absorption of the laser light is negligible at those temperatures.

\clearpage
\section{Estimation of mode temperature and force noise}
This Section discusses how the values and errors reported for cantilever mass, spring constant, mode temperature and force noise were calculated.

\subsection{Room temperature calibration}
\begin{figure}[ht]
\centering
	\includegraphics[width=10cm]{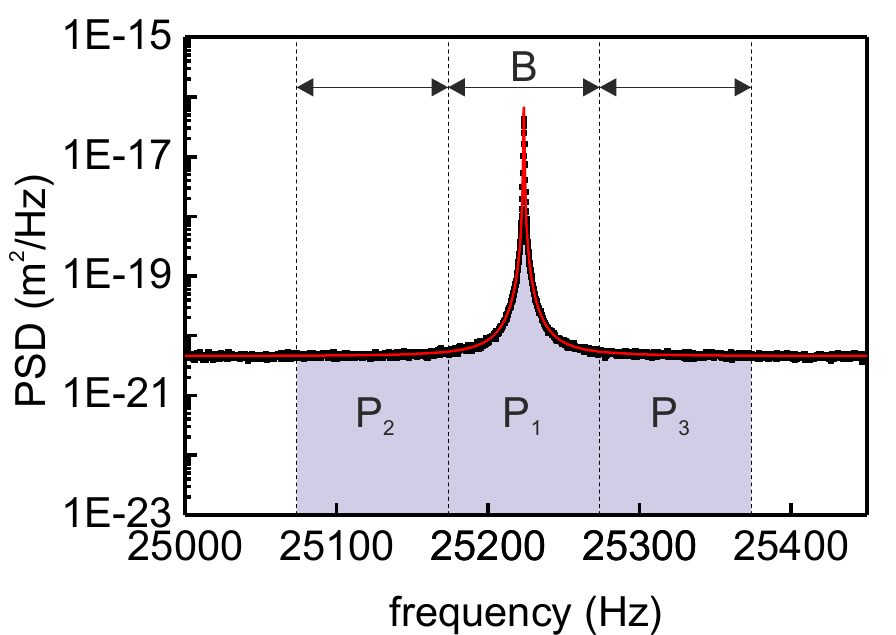}
	\caption{Integration of the power spectral density. All 3 powers $P_{1,2,3}$ are computed by integrating the power spectral density over the same interval $B$. The power under the peak is $P_1$ and $P_{2,3}$ are the baseline powers.  Data are for device \#1.}
	\label{fgr:PSDintegration}
\end{figure}

In a first step, the mass $m$ and spring constant $k$ of the resonator were determined by a room temperature measurement of the displacement power spectral density (PSD).  The room-temperature spring constant $\kRT$ and mass $m$ are given by
\begin{align}
\kRT &= \frac{\kB\TRT}{(\VrmsRT \sRT \cRT)^2}  \\
m &= \frac{\kRT}{(2\pi\fcRT)^2} 
\end{align}
where $\kB$ is Boltzmann's constant, $\TRT=295\unit{K}$, $\fcRT$ is the resonance frequency, $\VrmsRT$ the rms value of the displacement measured in units of Volts, $\sRT$ the interferometer displacement sensitivity in units of meters per Volt, and $\cRT$ the c-factor, a geometrical factor accounting for the mode shape at the location of the laser spot used for readout.  The rms displacement $\VrmsRT$ was calculated by integrating the displacement PSD.  $\VrmsRT$ corresponds to the area under the peak after the baseline has been substracted (see Fig. \ref{fgr:PSDintegration}).  To determine $\VrmsRT$, the voltage PSD $\SV(f)$ was numerically integrated around the resonance with a bandwidth $B$ to get the power $P_1$,
\begin{align}
P_1 &= \sum_{f_c-B/2}^{f_c+B/2} \SV(f) \Df
\end{align}
where the sum runs over $N$ points in the interval $[\fc-B/2,\fc+B/2]$ and $\Df = B/N$ is the frequency sampling of the PSD. To determine the baseline, two additional noise powers were calculated,
\begin{align}
P_2 &= \sum_{f_c-3B/2}^{f_c-B/2} \SV(f) \Df \\
P_3 &= \sum_{f_c+B/2}^{f_c+3B/2} \SV(f) \Df
\end{align}
corresponding to the areas to the left and right of the signal peak, respectively (see Fig. \ref{fgr:PSDintegration}).  The rms value of the displacement is then
\begin{align}
\Vrms &= \sqrt{P_1 - (P_2+P_3)/2}
\end{align}
The displacement sensitivity $\sRT$ was determined by sweeping the wavelength of the interferometer (by sweeping the temperature of the laser diode), and reading off the average slope from the interferometer fringe pattern.
The c-factor was computed based on the location of the reflective paddle and was $c=1.07$ for diamond and $c=1.04$ for silicon at room temperature.

\subsection{Errors of spring constant and mass}
The dominant contribution to the errors in the estimates for spring constant and mass is the uncertainty in the displacement sensitivity $\sRT$.  Because of the low reflection of the mirror paddle and due to additional stray reflections in our experimental setup, the displacement sensitivity could vary by about 15\% between different measurements.  We therefore assume an error $\sigma_{\sRT} = 0.15\sRT$  for the following error calculation.  The errors in the c-factor $\cRT$ and the rms displacement $\Vrms$ were comparably much smaller, typically $\sigma_{\cRT}<0.01\cRT$ and $\sigma_{\VrmsRT}<5\cdot10^{-4}\VrmsRT$, and are neglected in the following.  The propagated error is
\begin{align}
\begin{split}
	\sigma_{\kRT}
	&\approx  \frac{2 \kB \TRT}{\sRT^3 (\VrmsRT \cRT)^2} \sigma_{\sRT}
\end{split}
\label{equ:errm}
\end{align}
The corresponding error for the mass is
\begin{align}
\sigma_m &= \frac{\sigma_{\kRT}}{(2\pi\fcRT)^2}
\end{align}

\subsection{Low-temperature measurement}

To determine the mode temperature and thermal force noise at millikelvin temperatures, we used the room-temperature calibration of $\kRT$ and $m$ together with a low-temperature measurement of the displacement PSD.  The mode temperature $\TmK$ and thermal force noise PSD $\SFmK$ are given by
\begin{align}
\TmK &= \frac{\kmK(\VrmsmK\smK\cmK)^2}{\kB} \\
\SFmK &= \left[ \frac{4\kmK^2(\VrmsmK\smK\cmK)^2}{(2\pi\fcmK)\QmK} \right]^{1/2}
\end{align}
where $\kmK = \kRT (\fcmK/\fcRT)^2$ is the low-temperature spring constant, and $\VrmsmK$, $\smK$ and $\cmK$ are the corresponding displacement noise, displacement sensitivity and c-factor for the low-temperature measurement. $\QmK$ is the low-temperature quality factor that was determined in a separate ring-down measurement.

The principal uncertainties in $\TmK$ and $\SFmK$ are the calibration uncertainty in $\kRT$ and the uncertainty in the displacement sensitivity $\smK$.  Error propagation yields
\begin{align}
\sigma_{\TmK} &= \frac{1}{\kB} \left[ (\VrmsmK\smK\cmK)^4\sigma_{\kmK}^2 + (2\kmK\smK\VrmsmK^2\cmK^2)^2\sigma_{\smK}^2 \right]^{1/2} \\
\sigma_{\SFmK} &= \left[ \left(\frac{2(\VrmsmK\cmK\smK)^2}{\pi\fcmK\QmK}\right)\sigma_{k_{mK}}^2 + \left(\frac{2(\kmK\VrmsmK\cmK)^2}{\pi\fcmK\QmK}\right)\sigma_{\smK}^2 \right]^{1/2}
\end{align}

\subsection{Monte Carlo-based error propagation}
Because of the significant uncertainty in the calibration of the displacement sensitivities $\sRT$ and $\smK$, we performed a Monte Carlo simulation of the uncertainty in the force noise in addition to a standard error propagation calculation.  We found that the Monte Carlo method gives slightly different, but similar values to the standard error propagation.  Errors reported in the manuscript represent the values of the Monte Carlo simulation.

\subsection{Thermal force noise vs. force sensitivity}
The total noise power spectral density (PSD) measured at the output of the interferometer is the sum of the transduced displacement noise PSD of the cantilever $S_{v,th}$ (caused by the thermal force noise) and the detector noise PSD intrinsic to the interferometer $S_{v,det}$,
\begin{align}
S_{v,tot} = S_{v,th}G^2 + S_{v,det}.
\end{align}
Here, we use $G=\smK^{-1}$ ($G=\sRT^{-1}$ at room temperature) for the total transducer gain of the interferometer in units of V/m and the units of $S_{v,th}$ and $S_{v,det}$ are m$^2$/Hz and V$^2$/Hz, respectively. In order to be detectable, a force signal must cause a displacement signal that is larger than the standard deviation of the total noise measured with a certain bandwidth $B$, which according to Parseval's theorem is given by 
\begin{align}
\int_B S_{v,tot}(f) df.
\end{align}
In the narrow-filter limit ($B\ll f_0/Q$) and defining the force sensitivity as the smallest measurable force which can be detected with unit signal-to-noise ratio per bandwidth, $\Fmin$, we get
\begin{align}
\Fmin = \left[ B(S_{F,th} + \frac{k^2}{Q^2}\frac{S_{v,det}}{G^2}) \right]^{1/2}
\end{align}
where $S_{F,th} = 4 k_B T\gamma$ is the white thermal force noise PSD and $S_{v,det}$ is the displacement noise PSD of the interferometer determined by the baseline in Fig. \ref{fgr:PSDintegration} (note that $S_{v,det}$ does not correspond to a real cantilever displacement but to a displacement uncertainty caused by the detector noise). Since $S_{F,th}$ generates a peak of magnitude $S_{v,th,peak} = G^2 (Q/k)^2 S_{F,th}$ in the real displacement PSD, the force sensitivity can be expressed as
\begin{align}
\Fmin = \Fth \left[1 + \frac{S_{v,det}}{S_{v,th,peak}}\right]^{1/2}
\end{align}
where we have used the notation $\Fth = \sqrt{S_{F,th} B}$ for the force noise. Thus, in the absence of significant detector noise ($S_{v,det}\ll S_{v,th,peak}$), force noise and force sensitivity are equal, $\Fmin\approx\Fth$.  In the presence of detector noise, the force sensitivity is worse than the thermal force noise by a factor of $[1 + \frac{S_{v,det}}{S_{v,th,peak}}]^{1/2}$.

\section*{References}
\begin{itemize}
\item [{[S1]}] 
Y.~Tao and C.~Degen, ``{Facile Fabrication of Single-Crystal-Diamond
  Nanostructures with Ultrahigh Aspect Ratio},'' {\em Advanced Mater.},
  vol.~25, pp.~3962--3967, 2013.

\item [{[S2]}] 
Y.~Tao, J.~Boss, B.~Moores, and C.~Degen, ``Single-crystal diamond
  nanomechanical resonators with quality factors exceeding one million,'' {\em
  Nature Communications}, vol.~5, p.~3638, 2014.

\item [{[S3]}] 
R.~Hill, J.~Notte, and B.~Ward, ``The alis he ion source and its application to
  high resolution microscopy,'' {\em Physics Procedia}, vol.~1, pp.~135--141,
  2008.

\item [{[S4]}]  
O.~Scholder, K.~Jefimovs, I.~Shorubalko, C.~Hafner, U.~Sennhauser, and G.~L.
  Bona, ``Helium focused ion beam fabricated plasmonic antennas with sub-5 nm
  gaps,'' {\em Nanotechnology}, vol.~24, no.~395301, 2013.

\item [{[S5]}]  
Y.~Tao, P.~Navaretti, R.~Hauert, U.~Grob, M.~Poggio, and C.~L. Degen,
  ``Permanent reduction of dissipation in nanomechanical si resonators by
  chemical surface protection,'' {\em Nanotechnology}, vol.~26, pp.~465501--,
   2015.

\end{itemize}

\end{document}